\documentclass[%
 reprint,
superscriptaddress,
showpacs,preprintnumbers,
bibnotes,
 amsmath,amssymb,
 aps,
]{revtex4-1}

\usepackage{graphicx}% Include figure files
\usepackage{dcolumn}% Align table columns on decimal point
\usepackage{bm}% bold math
\usepackage[dvips]{color}

\begin{document}

%\preprint{APS/123-QED}

\title{Ruderman-Kittel-Kasuya-Yosida interaction in Weyl semimetals}% Force line breaks with \\
\author{Mir Vahid Hosseini}
\email{mv.hosseini@znu.ac.ir}
\affiliation{Department of Physics, Faculty of Science, University of Zanjan, Zanjan 45371-38791, Iran}
%\author{-------------}
%\affiliation{------------------, ---------------------, Iran}
\author{Mehdi Askari}
\affiliation{Department of Physics, Faculty of Science, Salman Farsi University of Kazerun, Kazerun, Iran}
\date{\today}% It is always \today, today,

\begin{abstract}
We theoretically demonstrate the Ruderman-Kittel-Kasuya-Yosida interaction between magnetic impurities that is mediated by the Weyl fermions embedded inside a three-dimensional Weyl semimetal (WSM). The WSM is characterized by a pair of Weyl points separated in the momentum space. Using the Green's function method and a two-band model, we show that four terms contribute to the magnetic impurity interaction in the WSM phase: the Heisenberg, Dzyaloshinsky-Moriya, spin-frustrated and Ising terms. Except the last term which is vanishingly small in the plane perpendicular to the line connecting two Weyl points, all the other interaction terms are finite. Furthermore, the magnetic spins of the Dzyaloshinsky-Moriya and spin-frustrated terms lie in the plane perpendicular to the line connecting two Weyl points, but in this plane, the magnetic spins of the Ising term have no components. For each contribution, an analytical expression is obtained, falling off with a spatial dependence as $R^{-5}$ at Weyl points and showing beating behavior that depends on the direction between two magnetic impurities.
\end{abstract}


\maketitle
\section{Introduction}
New classes of materials with extraordinary band structure including graphene \cite{graphene}, topological insulators \cite{TI}, and Weyl semimetals (WSMs) \cite{WeylSM1,WeylSM2,WeylSM3,WeylSM4} have attracted a great deal of interest both from the fundamental point of view and potential applications. Recently, some experiments revealed WSM phase in several chemical compounds \cite{Exp1,Exp2,Exp3,Exp4,Exp5,Exp7,Exp8} and photonic crystals \cite{phoExp1,phoExp2}. In the context of optical lattices, some schemes to realize WSM phase have also been proposed \cite{opl1,opl2,opl3,opl4}. WSMs have a novel nontrivial topological electronic states in their band structure. They possess gapless bulk and Fermi arc surface states, in contrast to the topological insulators which have gapless surface state inside the bulk gapped states. In WSMs which are three-dimensional (3D) extension of graphene's low-energy spectrum, carriers resemble the Weyl fermions around the so-called Weyl point where the conduction and valence bands touch each other.

In the absence of either time-reversal or parity symmetry, Dirac points split into pairs of Weyl points in the Brillouin zone with opposite chirality \cite{MagTIWeylSur} in WSMs. Time-reversal breaking case can be implemented by magnetic doping either in the bulk \cite{MagTIWeylBulk} or on the surface \cite{MagTIWeylSur} of topological insulators. Inversion symmetry breaking can be achieved by using an electric field \cite{ElecWeyl}, staggered strain \cite{strainWeyl}, alloying or application of external pressure \cite{PressWeyl1,PressWeyl2,PressWeyl3}. Remarkably, recently, the WSM has been studied in the nonmagnetic materials TaAs \cite{Exp7,Exp8,TheoTaAs1,TheoTaAs2}, NbAs \cite{ExpNbAs} and TaP \cite{ExpTaP1,ExpTaP2} without breaking of time-reversal symmetry.
Different non-doubly degenerate stable chiral states along with 3D relativistic nature of WSMs make a good playground to reach exotic phenomena such as the chiral anomaly \cite{chAnoma0,chAnoma1,chAnoma7,chAnoma8}, anomalous Hall effects \cite{chAnoma1,QHall}, and unconventional superconductivity \cite{uncSuper3,uncSuper7}.

Due to recent development of new materials with non-trivial topology, the investigation of indirect exchange interactions between magnetic impurities through carriers of a host material, known as the Ruderman-Kittel-Kasuya-Yosida (RKKY) interaction~\cite{RKKYPertu1,RKKYPertu2,RKKYPertu3}, has become an interesting issue. Non-trivial electronic states of carriers along with dimensions of space containing itinerant carriers result in an interesting behavior of the RKKY interaction between magnetic spins in graphene \cite{RKKYVozmediano,RKKYDugaev,saremi,RKKYSherafati1,RKKYSherafati2,RKKYBrey,RKKYBunder,RKKYSchaffer,Ferreira1,Klinovaja,RKKYBilayer1,RKKYBilayer2,RKKYBilayer3}, topological insulators \cite{RKKYTI1,RKKYTI2,RKKYTI3,RKKYTI4,RKKYTI5}, transition-metal dichalcogenides \cite{AsgariMoS2,dichalcogenides} and silicene \cite{RKKYsilicene1,RKKYsilicene2}.
The aforementioned features of WSMs would provide an important effect on the interaction of dilute impurities located inside these materials with the itinerant Weyl fermions. In this context, it is interesting to know how 3D feature and non-doubly degenerate Dirac dispersion relation emerging from the band structure, may affect the indirect exchange interaction between magnetic impurities in the WSM. In previous works, most of the studies have been focused on the ordering of host magnetic components \cite{ChiralRKKY0,ChiralRKKY1,ChiralRKKY2}, Kondo effect \cite{Kondo,Kondo1} of dilute magnetic impurities and the influence of surface \cite{FridelSur} or bulk \cite{FridelBul} states on density responses due to non-magnetic impurities in the WSM phase. However, the RKKY interaction between magnetic impurities in the WSM has remained unexplored so far.

In this paper, using the Green's function method and employing an effective two-band model, we address bulk properties of the RKKY interaction between two magnetic impurities in the WSM that is characterized by a pair of Weyl points separated in momentum space. It is found that four terms, which include the Heisenberg, Dzyaloshinsky-Moriya, spin-frustrated, and Ising couplings, contribute to magnetic spin interactions. Impurity spin components of the Dzyaloshinsky-Moriya and spin-frustrated interactions lie in the plane perpendicular to the direction of the line joining the two Weyl points, but the Ising interaction has only spin components in the direction parallel to the line connecting two Weyl points. While the range function of the Ising interaction vanishes in the direction perpendicular to the line connecting the Weyl points, the range functions of other terms survive in this direction. Furthermore, we also demonstrate that the spatial dependence of the range functions fall off as $R^{-5}$ $(R^{-3})$ for zero (finite) Fermi energy and show a beating behavior depending on the direction between two magnetic impurities.

%%%%%%%%%%%%%%%%%%%%%%%%%%%%%%%%%%%%%%%%%%%%%%%%%%%%%%%%%%%%%%%%%%%%%%%
\begin{figure}[t]%[!htb]
\begin{center}
\includegraphics[width=5.8cm]{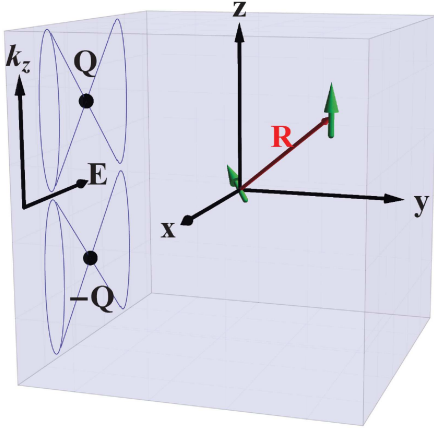}
\caption{(Color online) Schematic of WSM including two magnetic impurities (green arrows). One of the magnetic impurities is located at the origin and the other one is separated from the origin by a vector $\textbf{R}$. The figure also shows linearly dispersing excitation spectrum around a pair of Weyl points separated by a distance 2$Q$ in the $z$-direction of momentum space.}
\label{WSM}
\end{center}
\end{figure}
%%%%%%%%%%%%%%%%%%%%%%%%%%%%%%%%%%%%%%%%%%%%%%%%%%%%%%%%%%%%%%%%%%%%%%%%

\section{Model and Theory}\label{sec2}
We consider a 3D WSM with a pair of Weyl points separated in the Brillouin zone containing two magnetic impurities, as shown schematically in Fig. \ref{WSM}. The strength of the magnetic impurities is supposed to be weak enough to keep linearly dispersing bands \cite{impurity1,impurity2,impurity3}. The effective Hamiltonian describing the Weyl fermions is given by \cite{chAnoma1,chAnoma7,uncSuper3},
\begin{eqnarray}
H_0 &=& \sum_{\mathbf q,\tau} \psi^\dagger_{\tau}(\mathbf q) h_{\tau}(\mathbf q) \psi_{\tau}(\mathbf q),
\end{eqnarray}
with
\begin{eqnarray}
h_{\tau}(\mathbf q) &=& v_0(q_x\sigma_x+q_y\sigma_y -\tau q_z\sigma_z),
\end{eqnarray}
where $\psi_{\tau}(\mathbf q) = (\psi_{\tau\uparrow}(\mathbf q), \psi_{\tau\downarrow}(\mathbf q))^T$ is the Weyl spinor, $\tau = \pm$ represents the pair of Weyl points with opposite chirality which are located at $\mathbf{P}_{\tau} = (0, 0, \tau Q)$ in momentum space, $v_0$ is the Fermi velocity, and $\sigma_{x,y,z}$ are the Pauli matrices for the spin degree of freedom.

We assume that the magnetic impurities are embedded inside the WSM such that the effect of Fermi arc surface states can be neglected. As a result of this assumption, the bulk states provide the main contribution to the indirect exchange interaction. The interaction of the spin of magnetic impurities with the Weyl fermions of host material can be described by a contact interaction,
\begin{equation}
H_{\text{int}} = J \sum_{j=1,2} \mathbf{S}_j \cdot \mathbf{s}(\mathbf{R}_j),
\end{equation}
where $\mathbf  s(\mathbf{r}) = \frac{1}{2} \sum_i \delta(\mathbf{r} - \mathbf{r}_i)\boldsymbol \sigma_i$
indicates the spin density for Weyl fermion at position coordinate $\mathbf{r}$ ($\hbar = 1$), and $\mathbf{S}_j$ is the
localized impurity spin at site $\mathbf{R}_j$. For simplicity and concreteness, here, we suppose that the exchange coupling $J$ is isotropic for each component of impurities spin and homogeneous throughout the WSM and also, the first impurity is situated on the origin $\mathbf R_1$ = $(0, 0, 0)$.

Treating $H_{\text{int}}$ as a perturbation to $H_0$, at second order of perturbation, one can obtain an effective interaction between the localized spins of magnetic impurities~\cite{RKKYPertu1,RKKYPertu2,RKKYPertu3,RKKYPertu4},
\begin{equation}
\begin{split}
&H_{\text{RKKY}} = -\frac{J^2}{\pi}\times\\
 &Tr \left[ \int_{-\infty}^{\epsilon_F} \!\!d\epsilon\:
 \Im\left\{(\mathbf{S_1}\cdot\mathbf{\sigma})  G^0(\mathbf R,\epsilon^+) (\mathbf{S_2}\cdot\mathbf{\sigma}) G^0(-\mathbf R,\epsilon^+)
\right\}\right],
\label{HRKKY}
\end{split}
\end{equation}
where $\epsilon^+=\epsilon+i 0^+$, $Tr$ means trace over spin space, $\Im$ is imaginary part, $\epsilon_F$ is the Fermi energy measured from the Weyl point, $\mathbf{R} = \mathbf R_2 - \mathbf R_1$ is the vector connecting the two magnetic centers and $G^0(\mathbf R,\epsilon^+)$ stands for the $2\times 2$ Green's function matrix in real-space which around the two Weyl points can be expressed by,
\begin{eqnarray}
G^0 (\mathbf{R}, \epsilon^+)& = &\int \frac{d^3q}{\Omega_{BZ}} \ e^{i \mathbf{q} \cdot \mathbf{R}}\sum_{\tau}e^{i \mathbf{P}_{\tau} \cdot \mathbf{R}} G^0_{\tau} (\mathbf{q}, \epsilon^+),
\label{G-RE}
\end{eqnarray}
with
\begin{equation}
G^0_{\tau} (\mathbf{q}, \epsilon^+) = [\epsilon^+ - h_{\tau}(\mathbf q)]^{-1},
\label{G-MO}
\end{equation}
where $G^0_{\tau} (\mathbf{q}, \epsilon^+)$ is the momentum space Green's function and $\Omega_{BZ}$ is the area of the first Brillouin zone.
In Eq. (\ref{G-RE}), the exponential factor $e^{i \mathbf{q} \cdot \mathbf{R}}$ can be expanded in terms of spherical harmonics according to the Rayleigh equation \cite{Rayleigh},
\begin{eqnarray}
e^{i \mathbf{q} \cdot \mathbf{R}} = 4\pi \sum^{\infty}_{l=0}\sum^{l}_{m=-l}\!i^lj_l(qR)Y^{\ast}_{lm}(\theta_R,\phi_R)Y_{lm}(\theta_q,\phi_q),
\label{Rayleigh}
\end{eqnarray}
where $j_l$ is the spherical Bessel function of order $l$, $Y_{lm}$ is the spherical harmonic function, $(R, \theta_R,\phi_R)$ and $(q, \theta_q,\phi_q)$ are the spherical coordinates of $\mathbf{R}$ and $\mathbf{q}$, respectively.

Upon substituting Eq. (\ref{Rayleigh}) into Eq. (\ref{G-RE}), we find analytical expression for the real-space Green's function as,
\begin{eqnarray}
G^0 (\pm\mathbf{R}, \epsilon^+)& = &-\frac{4\pi^2}{\Omega_{BZ}}\cos(Qz)\frac{e^{\frac{iR\epsilon^+}{v_0}}}{Rv_0^2}\{\epsilon^+ \sigma_0+\frac{R\epsilon^++iv_0}{R}\nonumber\\
&\times&(\pm\hat{\rho}-iW\hat{z})\cdot \mathbf{\sigma}\},
\label{Green}
\end{eqnarray}
where $\sigma_0$ is the unit matrix, $z = R \cos(\theta_R)$, $W = \tan(Qz)\cos(\theta_R)$, and the unit vector orthogonal to the line connecting the two Weyl points is $\hat{\rho} = \hat{R} - \cos(\theta_R)\hat{z}$ with $\hat{R}$ and $\hat{z}$ being the unit vectors along the $\mathbf{R}$ and the $z$-axis, respectively.
Since we have used effective Hamiltonian, unphysical divergence would occur in integrating over the occupied states of the valance band (states between $\epsilon = -\infty$ and zero) in Eq. (\ref{HRKKY}). In order to avoid this problem, one can use either the cut-off procedure \cite{saremi} or Matsubara Green's functions in the coordinate-imaginary time representation \cite{kogan}. Following Ref. \cite{saremi}, we multiply the integrand by a smooth cutoff function.
Inserting Eq. (\ref{Green}) into Eq. (\ref{HRKKY}) and taking into account the point above, we arrive at the final expression for $H_{RKKY}$ [Eq. (\ref{HRKKY})] as
\begin{eqnarray}
 H_{\text{RKKY}} &=& F_1 \mathbf S_1 \cdot \mathbf S_2 + F_2 \hat{\rho} \cdot (\mathbf S_1 \times \mathbf S_2)\nonumber\\
 &+& F_3 (\mathbf S_1 \cdot \hat{\rho})(\mathbf S_2 \cdot \hat{\rho}) + F_4 \mathbf{S}^z_1 \mathbf{S}^z_2,
\label{finalHRKKY}
\end{eqnarray}
where the range functions are
\begin{eqnarray}
F_1 &=& -\frac{8\pi^3J^2}{\Omega^2_{BZ}v_0R^5}I_1,\label{rangeF1}\\
F_2 &=& \frac{64\pi^3J^2}{\Omega^2_{BZ}v_0R^5}I_2,\label{rangeF2}\\
F_3 &=& \frac{16\pi^3J^2}{\Omega^2_{BZ}v_0R^5}I_3,\label{rangeF3}\\
F_4 &=& \frac{16\pi^3J^2}{\Omega^2_{BZ}v_0R^5}I_4.
\label{rangeF4}
\end{eqnarray}
Here, we have defined the dimensionless couplings, $I_{1}$, $I_{2}$, $I_{3}$ and $I_{4}$ as
\begin{eqnarray}
I_1 &=& \cos^2(Qz)[(1-2\gamma^2)\cos(2\gamma)+2\gamma\sin(2\gamma) \nonumber\\
 &+& (\sin^2(\theta_R)+W^2)I_0],\label{DLrange1}\\
I_2 &=& \cos^2(Qz)[\gamma\cos(2\gamma)+(\gamma^2-1)\cos(\gamma)\sin(\gamma)],\label{DLrange2}\\
I_3 &=& \cos^2(Qz)I_0,\label{DLrange3}\\
I_4 &=& \sin^2(Qz)\cos^2(\theta_R)I_0,
\label{DLrange4}
\end{eqnarray}
with $I_0 = (5-2\gamma^2)\cos(2\gamma)+6\gamma\sin(2\gamma)$, and $\gamma = \frac{R\epsilon_F}{v_0}$. We note that there is no dependence on the azimuthal angle $\phi_R$ in Eqs. (\ref{DLrange1})-(\ref{DLrange4}). This arises as
a consequence of the azimuthal symmetry around a pair of Weyl dispersions.

The resulting RKKY interaction [see Eq. (\ref{finalHRKKY})] consists of four different terms: the Heisenberg term, the Dzyaloshinsky-Moriya term, the spin-frustrated term and the Ising term, whose range functions are $F_1$, $F_2$, $F_3$ and $F_4$, respectively. Notice that magnetic spins of both the Dzyaloshinsky-Moriya and the spin-frustrated terms have no components in the direction parallel to the line connecting the Weyl points. Therefore, the Dzyaloshinsky-Moriya interaction favor twisted impurity spin structure in the plane perpendicular to the line connecting the Weyl points, whereas the spin-frustrated interaction causes (anti) parallel alignment of the impurity spins along the projection of the line joining the two magnetic impurities on this plane. The Ising term describes antiferromagnetic or ferromagnetic ordering, depending on its sign, between the components of impurity spins pointing from one Weyl point to the other one. While both the spin-frustrated and the Ising terms have the same prefactors with positive sign, the Heisenberg and the Dzyaloshinsky-Moria terms have the smallest and the largest perfactors with negative and positive signs, respectively.

At Weyl points, $\epsilon_F = 0$, all the range functions [Eqs. (\ref{rangeF1}) - (\ref{rangeF4})] show a decaying behavior with the spatial dependence as $R^{-5}$ similar to that previously found in carbon nanotubes with center-adsorbed impurities \cite{RKKYNanotubes}. This behavior can be described based on semiclassical arguments \cite{semiclassicRKKY} which for any d-dimensional material with density of states $N(\epsilon) \propto |\epsilon|^{\alpha}$ at Fermi energy, the RKKY interaction decays at large distances with a power law decay $R^{(-d-\alpha)}$. Since, for a 3D WSM ($d = 3$) at Weyl points, the density of states decreases with a power $\alpha = 2$, indeed, one can obtain the power law decay $R^{-5}$. On the other hand, for finite Fermi energy and at large distances, $R \gg 1$, the range functions can be approximated by,
\begin{eqnarray}
F_1 &\approx& \frac{16\pi^3J^2\epsilon^2_F\cos^2(2\gamma)}{\Omega^2_{BZ}v^3_0R^3}\cos^2(Qz)[\sin^2(\theta_R)\!+\!W^2],\\
F_2 &\approx& \frac{32\pi^3J^2\epsilon^2_F\cos^2(2\gamma)}{\Omega^2_{BZ}v^3_0R^3}\cos^2(Qz),\\
F_3 &\approx& -\frac{32\pi^3J^2\epsilon^2_F\cos^2(2\gamma)}{\Omega^2_{BZ}v^3_0R^3}\cos^2(Qz),\\
F_4 &\approx& -\frac{32\pi^3J^2\epsilon^2_F\cos^2(2\gamma)}{\Omega^2_{BZ}v^3_0R^3}\sin^2(Qz)\cos^2(\theta_R),
\end{eqnarray}
which show the power law decay $R^{-3}$. Consequently, in the WSM phase, the range functions are rather short-range and fall off faster than those of the doped (undoped) graphene \cite{saremi}, doped (undoped) topological insulators \cite{RKKYTI1}, 2D electron gas systems \cite{RKKYPertu1,RKKYPertu2,RKKYPertu3} and monolayer transition-metal dichalcogenides \cite{dichalcogenides} which slowly decay as $R^{-2}$ ($R^{-3}$).

Moreover, due to momenta-shift of the pair of Weyl points in the WSM phase, the dimensionless couplings $I$'s depend on the direction between two magnetic impurities which is similar to the results obtained in graphene \cite{RKKYSherafati1,RKKYSherafati2}, monolayer transition-metal dichalcogenides \cite{AsgariMoS2} and silicene \cite{RKKYsilicene1}. However, there is a key difference from these materials. In graphene, the direction dependence of couplings is identical up to an additional phase factor \cite{RKKYSherafati2}, also the direction dependence of all couplings is not identical in monolayer transition-metal dichalcogenides \cite{AsgariMoS2} and silicene \cite{RKKYsilicene1}. But, obviously, in the WSM case, the dependence of the dimensionless couplings $I_2$ and $I_3$ on direction is the same, while $I_1$ and $I_4$ have different direction dependence which can be associated to the opposite chirality of the pair of Weyl points along with 3D nature of WSM. Interestingly, if both impurities place on the plane perpendicular to the line connecting two Weyl points, $\theta_R = \pi/2$, the range function of Ising term vanishes, $F_4 = 0$. Also, for undoped case, $\epsilon_F = 0$, unlike the other couplings, the Dzyaloshinsky-Moria term vanishes \cite{RKKY12DSpinOrbit} regardless of direction and distance between impurities.

%%%%%%%%%%%%%%%%%%%%%%%%%%%%%%%%%%%%%%%%%%%%%%%%%%%%%%%%%%%%%%%%%%%%%%%
\begin{figure}[t]
\begin{center}
\includegraphics[width=7cm]{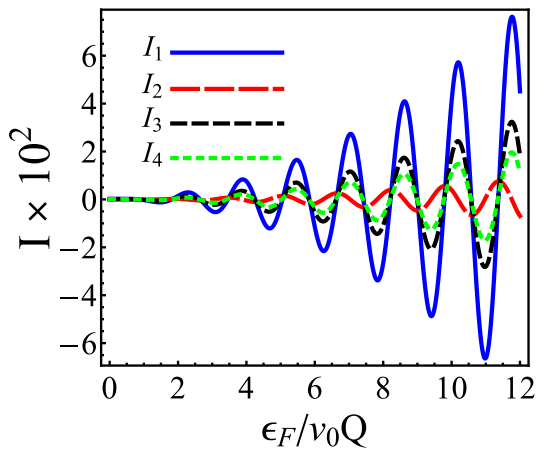}
\caption{(Color online) Dimensionless couplings $I$'s as a function of the dimensionless Fermi energy ($\epsilon_F/v_0Q$) for $\theta_R = \pi/3$ and $QR = 2$.}
\label{IFerTeta}
\end{center}
\end{figure}
%%%%%%%%%%%%%%%%%%%%%%%%%%%%%%%%%%%%%%%%%%%%%%%%%%%%%%%%%%%%%%%%%%%%%%%%

\section{Discussion of the Results}\label{sec3}
In our considerations, we choose $Q^{-1}$ as the length unit and $v_0Q$ as the energy unit. In Fig. \ref{IFerTeta}, the dependence of the dimensionless couplings $I$'s on the Fermi energy is shown for $\theta_R = \pi/3$ and $QR = 2$. One can see that all the $I$'s oscillate and their amplitudes increase with increasing dimensionless Fermi energy $\epsilon_F/v_0Q$, because of increasing the Fermi surface of the system. Furthermore, we observe clearly that while $I_1$, $I_3$ and $I_4$ in terms of $\epsilon_F$ are in-phase, $I_2$ has a phase shift with respect to the others.

\begin{figure}[t]
\begin{center}
\centering
\includegraphics[width=7cm]{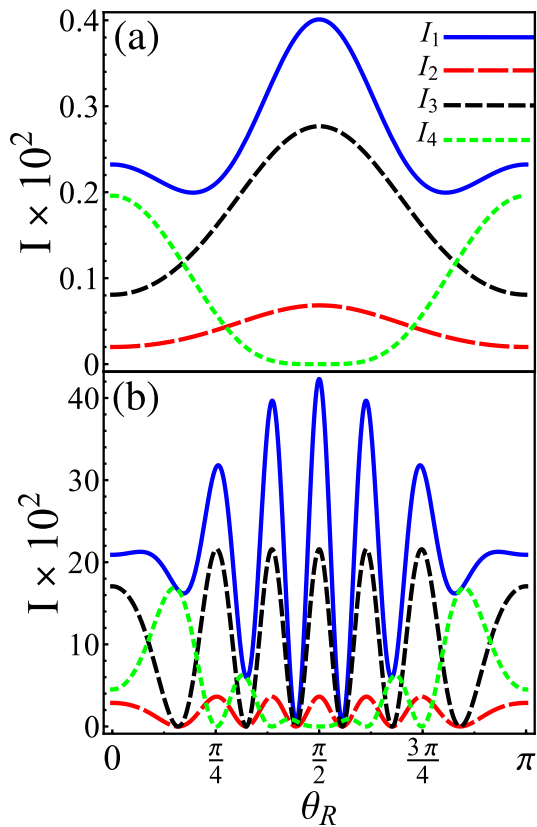}
\caption{(Color online) Dimensionless couplings $I$'s as a function of the direction between two magnetic impurities $\theta_R$ for (a) $QR = 1$, (b) $QR = 8.95$ with $\epsilon_F = 4 v_0Q$.}
\label{ITeta}
\end{center}
\end{figure}

The dimensionless couplings $I$'s versus $\theta_R$ strongly modulate as depicted in Fig. \ref{ITeta}. Panels (a) and (b) refer to the cases $QR = 1$ and $QR = 8.95$, respectively, with $\epsilon_F = 4 v_0Q$ indicating that the number of oscillations depend on the value of dimensionless distance $QR$ and increase at large value of $QR$. Moreover, from Eqs. (\ref{DLrange1}) - (\ref{DLrange4}), it is easy to see that the maxima of these oscillations for the dimensionless couplings $I_1$, $I_2$ and $I_3$ ($I_4$) take place at certain angles $\theta_R \simeq \arccos[n\pi/(QR)]$ ($\arccos[(2n-1)\pi/(2QR)]$) with $n = 0, \pm 1, \pm 2, \ldots$. We note that while $I_1$, $I_2$ and $I_3$ have finite values of amplitude modulation around the plane perpendicular to the line connecting two Weyl points, $\theta_R = \pi/2$, as mentioned above, the values of $I_4$ become vanishingly small around such angle. In addition, $I_1$ has the largest amplitude modulation around $\theta_R = \pi/2$.
%%%%%%%%%%%%%%%%%%%%%%%%%%%%%%%%%%%%%%%%%%%%%%%%%%%%%%%%%%%%%%%%%%%%%%%
%%%%%%%%%%%%%%%%%%%%%%%%%%%%%%%%%%%%%%%%%%%%%%%%%%%%%%%%%%%%%%%%%%%%%%%%
%%%%%%%%%%%%%%%%%%%%%%%%%%%%%%%%%%%%%%%%%%%%%%%%%%%%%%%%%%%%%%%%%%%%%%%
\begin{figure}[t]
%\begin{center}
\includegraphics[width=7.5cm]{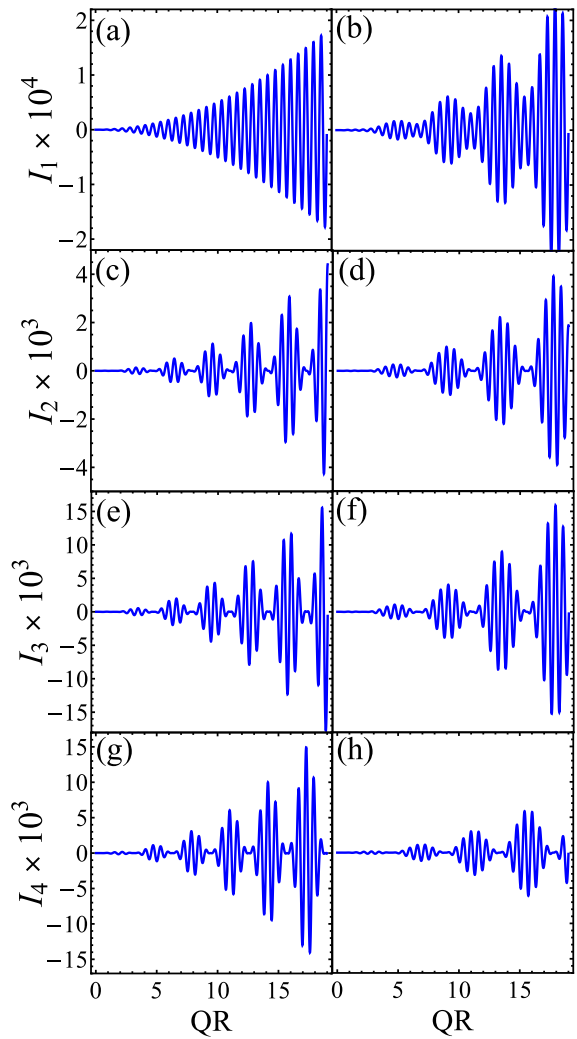}
\caption{(Color online) Dimensionless couplings (a)-(b) $I_1$, (c)-(d) $I_2$, (e)-(f) $I_3$  and (g)-(h) $I_4$ as functions of the dimensionless length $QR$. Left (right) panels are for $\theta_R = 0$ ($\pi/4$). Here $\epsilon_F = 5 v_0Q$.}
\label{IR}
%\end{center}
\end{figure}
%%%%%%%%%%%%%%%%%%%%%%%%%%%%%%%%%%%%%%%%%%%%%%%%%%%%%%%%%%%%%%%%%%%%%%%%

In Fig. \ref{IR}, we plot the dimensionless couplings $I$'s as functions of dimensionless distance $QR$ for $\theta_R = 0$ ($\pi/4$) [left (right) panels] with $\epsilon_F = 5 v_0Q$. Figure \ref{IR}(a) shows that $I_1$ increases oscillatory with respect to $QR$ in the $\theta_R = 0$ direction. But for $\theta_R = \pi/4$, $I_1$ exhibits a beating behavior as one can see in Fig. \ref{IR}(b). In this case, the upper and lower envelope functions increase oscillatory in magnitude without crossing points between them, indicating that superposed carriers' waves have different amplitudes. As illustrated in Figs. \ref{IR}(c) and (d) for $I_2$ (also Figs. \ref{IR}(e)-(f) for $I_3$ and Figs. \ref{IR}(g)-(h) for $I_4$), the dimensionless couplings almost always show a beating behavior as functions of dimensionless distance $QR$ with crossing points between the upper and lower envelope functions. It is interesting to
note that in all the cases discussed above the beat period increases as we increase $\theta_R$ from $0$ up to $\theta_R = \pi/2$. With further increase of the $\theta_R$, the situation reverses and the beat period decreases (not shown here), thus, as a consequence, the plane $\theta_R = \pi/2$ acts as a mirror plane.

Finally, it should be remarked that although we have considered indirect exchange interaction in the WSM phase with a pair of Weyl points separated in momentum space along the $z$-direction but our calculation can be extended directly to cases with multiple sets of Weyl nodes in the Brillouin zone with arbitrary direction between them as well.
Also, WSM phase has not been yet experimentally reported by breaking of time-reversal symmetry, but there are some materials such as TaAs \cite{Exp7,Exp8,TheoTaAs1,TheoTaAs2}, NbAs \cite{ExpNbAs} and TaP \cite{ExpTaP1,ExpTaP2} that have been recently found to exhibit 3D Weyl fermion states with time-reversal-symmetric feature. We believe that such materials are promising candidates to explore the RKKY interaction between magnetic impurities in WSMs.

\section{Conclusion}\label{sec4}
In this paper, we studied the RKKY interaction between magnetic impurities mediated by the bulk states of WSM within the two-band model and found that the 3D character of WSM with opposite chirality of Weyl points provides a unique feature which is absent in the 2D counterpart materials. We showed that the RKKY interaction can be decomposed as the Heisenberg, the Dzyaloshinsky-Moriya, the spin-frustrated and the Ising terms. For each term an analytic formula is derived that decays as the power law of $R^{-5}$ at Weyl points and shows the beating behavior depending on the impurities' orientations. The magnitude of the range functions remains finite in the plane perpendicular to the line connecting the two Weyl points, except for the case of the Ising interaction. Furthermore, the spins of magnetic impurities in the Dzyaloshinsky-Moriya and spin-frustrated interactions have no components in the direction parallel
to the line connecting two Weyl points, but the Ising interaction has only spin components in such direction.
\par
{\it Note added}-- After completion of the present paper we became aware that the same issue has been investigated in Ref. \cite{RKKYsem} and some results similar to our paper have been obtained.

\section*{Acknowledgment}

We would like to thank A. Ghorbanzadeh and A. Najafi for useful comments on this work. M.V.H is grateful to F. Mirzapour for his support.

\end{document}